\documentclass[11pt]{article}

\usepackage[a4paper,margin=1.8cm]{geometry}

\usepackage[T1]{fontenc}
\usepackage{lmodern}
\usepackage{microtype}

\usepackage{graphicx}
\usepackage{eso-pic}

\usepackage[hidelinks]{hyperref}

\usepackage{longtable}
\usepackage{xcolor}
\usepackage{array}
\usepackage{url}
\usepackage{booktabs}
\usepackage{amsfonts}
\usepackage{nicefrac}
\usepackage{subcaption}
\usepackage{float}
\usepackage[section]{placeins}
\usepackage[backend=biber,style=authoryear]{biblatex}
\graphicspath{ {./img/} }
\usepackage{enumitem}
\usepackage[symbol]{footmisc}

\date{March 2026}
  
\title{How people use Copilot for Health}

\author{%
    \begin{tabular}{c}
    \small
    Beatriz Costa-Gomes\footnotemark[1] \quad Pavel Tolmachev\footnotemark[1] \quad Eloise Taysom \quad Viknesh Sounderajah \\[0.3em]
    \small
    Hannah Richardson \quad Philipp Schoenegger \quad Xiaoxuan Liu \quad Matthew M Nour \\[0.3em]
    \small
    Seth Spielman \quad Samuel F.\ Way \quad Yash Shah \quad Michael Bhaskar \quad Harsha Nori \\[0.3em]
    \small
    Christopher Kelly \quad Peter Hames \quad Bay Gross \quad Mustafa Suleyman \quad Dominic King\\[0.6em]
    \normalsize Microsoft AI
    \end{tabular}
  }

\begin{document}

\AddToShipoutPictureBG*{%
  \AtPageUpperLeft{%
    \raisebox{-1.4cm}{\hspace{1.4cm}%
      \begin{minipage}{5cm}%
        \includegraphics[height=0.5cm]{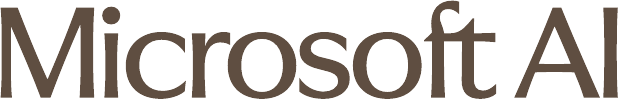}%
      \end{minipage}%
    }%
  }%
}

\maketitle

  \footnotetext[1]{Both BCG and PT contributed equally and have the right to list their name first in their CV. The co-first authorship order was determined via the best of five rounds in UNO.}
  
\begin{abstract}
We analyze over 500,000 de-identified health-related conversations with Microsoft Copilot from January 2026 to characterize what people ask conversational AI about health. We develop a hierarchical intent taxonomy of 12 primary categories using privacy-preserving LLM-based classification validated against expert human annotation, and apply LLM-driven topic-clustering for prevalent themes within each intent. Using this taxonomy, we characterize the intents and topics behind health queries, identify who these queries are about, and analyze how usage varies by device and time of day. Five findings stand out. First, nearly one in five conversations involve personal symptom assessment or condition discussion, and even the dominant general information category (40\%) is concentrated on specific treatments and conditions, suggesting that this is a lower bound on personal health intent. Second, one in seven of these personal health queries concern someone other than the user, such as a child, a parent, a partner, suggesting that conversational AI can be a caregiving tool, not just a personal one. Third, personal queries about symptoms and emotional health queries increase markedly in the evening and nighttime hours, when traditional healthcare is most limited. Fourth, usage diverges sharply by device: mobile concentrates on personal health concerns, while desktop is dominated by professional and academic work. Fifth, a substantial share of queries focuses on navigating healthcare systems such as finding providers, and understanding insurance, highlighting friction in the delivery of existing healthcare. These patterns have direct implications for platform-specific design, safety considerations, and the responsible development of health AI.

\end{abstract}


\section{Introduction}

Health is among the most high-stakes domains in which people interact with conversational AI.\footnotemark[2] For many users, conversational AI is becoming increasingly important in their medical journeys \parencite{huo2025large}, ranging from a first point of contact when a symptom appears, to medication questions and understanding interactions with healthcare professionals and the health system. The quality of these AI-powered medical interactions is thus central to individual wellbeing \parencite{goldberg2024no}, and understanding how these interactions currently play out is crucial for improving health-related conversational AI experiences.

\footnotetext[2]{Throughout this paper, we use ``AI'' as shorthand for large language model (LLM)-powered conversational AI applications, such as Microsoft Copilot, ChatGPT, and Gemini, unless otherwise specified. We recognize that AI encompasses a much broader set of technologies with diverse applications and risk profile, but these are outside the scope of this paper.}

Conversational AI is a stepwise change in how humans interact with information platforms and digital technology. Unlike web search, which returns ranked web pages for a single query, chat interfaces support multi-turn dialogue in which users can ask a question, add context, and correct course, producing responses tailored to their specific situation. In the early web era, \textcite{eysenbach2002} documented how ``Dr. Google'' reshaped patient-provider relationships, a trend later examined by \textcite{tan2017}. Conversational AI represents a step change in that trajectory, and is likely to fundamentally reshape how people approach their health in a digital context \parencite{topol2019}. 

A growing body of research has examined LLM capabilities in health contexts. Large language models perform competitively on medical licensing examinations \parencite{kung2023,nori2023capabilitiesgpt4medicalchallenge,nori2024medprompto1explorationruntime} and clinical reasoning benchmarks \parencite{Singhal2023,nori2025sequentialdiagnosislanguagemodels,brodeur2025superhumanperformancelargelanguage}. However, strong benchmark performance does not always translate to real-world reliability, as chatbots have been shown fail in identifying the severity of health issues in triage settings \parencite{ramaswamy2026chatgpt}, while users assisted by LLMs sometimes perform no better than controls at identifying conditions and choosing appropriate actions \parencite{bean2026reliability}. Despite these limitations, studies looking at patient-facing applications suggest users find AI-generated health responses comparable to or preferred over physician responses in some settings \parencite{lizee2024conversational, ruben2025artificial}. Crucially, it is often actionable guidance rather than mere information provision \parencite{lee2023} that has been empirically shown to drive engagement. Complementary work from Anthropic \parencite{anthropic2025} and OpenAI \parencite{openai2025} has documented affective and support-seeking interactions, indicating that conversational AI serves emotional as well as informational needs. Our previous work found that health-related queries on consumer Microsoft Copilot were the most prevalent topic category on mobile, a trend that remained consistent across temporal dimensions \parencite{costagomes2025}. 

However, despite this growing body of work, a basic question remains unanswered at scale: what, specifically, are people asking about regarding their health? We know that health is a dominant category of AI usage, but two dimensions remain uncharted: the \emph{intents} behind health queries -- the broad purpose of a conversation, such as seeking personalized coaching versus navigating the healthcare system -- and the \emph{topics} raised within each intent -- the specific subjects within that purpose, such as tailored meal plans or finding a local specialist.

This matters for three reasons. First, building an AI health experience that meets users where they are (on the right device, at the right time, and for the right person) requires knowing what they need. Second, ensuring that AI responds appropriately to different types of health queries depends on understanding the nature of the questions being asked. Copilot is not intended to replace professional medical advice, and it is important to know when to provide information versus when to direct users to professional care. Third, because conversational AI is a new modality for health information, these usage patterns are also likely to evolve. A baseline characterization is therefore essential, both for improving the experience now and for tracking how needs change over time. More broadly, as AI systems increasingly serve health-related needs, systematic characterization of these needs becomes essential for responsible development.

In this paper, we analyze over 500,000 health-related conversations from Microsoft Copilot to provide that baseline. We make three primary contributions: (1) We develop a hierarchical intent taxonomy comprising 12 primary categories and fine-grained topic clusters within each, using a mixed-methods approach validated against human annotation. (2) We apply this taxonomy to characterize health queries at scale, revealing the prevalence of personal health intents — including symptom assessment, condition management, and emotional wellbeing --- and showing that a substantial fraction of these serve caregiving rather than individual needs. (3) We analyze how intents vary by device and time of day, showing that context shapes the nature of health engagement with AI.

\section{Data \& Classification} 

\subsection{Privacy considerations}

All Copilot data used in this research was de-identified prior to analysis through a two-stage, privacy-preserving pipeline. In the first stage, raw conversation transcripts pass through an automated scrubbing process that detects and removes personally identifiable information (PII), including names, phone numbers, email and physical addresses, government‑issued identifiers such as social security and passport numbers, and financial details such as credit card and bank account numbers. In the second stage, an LLM generates a short, privacy‑preserving English‑language summary of each conversation that captures the topic and intent without reproducing the user's original words. All subsequent analysis, including our clustering, operates on these summaries rather than on any form of the original text. No human researcher accessed raw conversation content at any point, i.e., the entire pipeline follows an ``eyes‑off'' model in which only machine‑based classifiers interact with scrubbed data.

Data elements were limited to coarse, non‑identifying attributes and were used solely for aggregate analysis. No attempts were made to re‑identify users, infer individual health status, or draw conclusions about specific individuals.

All data processing occurred within Microsoft‑controlled systems with access controls and retention limits. Internal privacy review was conducted prior to and throughout the research. Data is retained only for the minimum period necessary to conduct the analysis and validate findings. All results are reported at an aggregate level. Copilot Health evaluation activities are not designed to generate generalizable knowledge, or evaluate clinical hypotheses. As such, these activities are not classified as human subjects research.

\subsection{Methodology}

This paper extends the methodology of the Copilot Usage Report 2025 \parencite{costagomes2025}. We sampled conversations from January 2026, excluding all enterprise, educational and commercial accounts. The sample is global, with around 22\% of conversations originating from the United States and the remainder from across the world; approximately 45\% of conversations are in English.

Each conversation is assigned a general topic (what the conversation is about), a general intent (what Copilot is expected to do), and a privacy-preserving English summary by the same pipeline described in the previous paper. From these, we draw on a random sample of over 500,000 conversations that have been classified as ``Health and Fitness'' as a general topic. 

For this subset, we apply a second classifier that assigns a health-specific intent from a 12-category taxonomy (see Table \ref{tab:intent}) and extracts structured attributes: including who the health query is about, and any symptoms or conditions explicitly mentioned by the user.

The taxonomy was developed prior to this study by in-house clinician scientists, informed by previous observational analyses of user intents. The intents of ``Other Health/Fitness'' and ``Not Health'' are used for telemetry purposes and will not be further explored in this paper. To validate the health-intent classifier, clinical scientists independently labeled a sample of conversations that were available for human review, showing agreement between the LLM classifier and human annotators.

To move from individual labels to the thematic categories reported in this paper, we employ an LLM-driven clustering method following TnT-LLM \parencite{wan2024}. For each health intent category, we provide an LLM with batches of conversation summaries and their extracted attributes, and prompt it to group conversations by the user's underlying journey, producing named clusters that describe coherent patterns (e.g., ``General Wellness, Food Choice, and Product Comparison'', ``Strength Training and Fitness Routine Planning'', ``Finding a Local Healthcare Provider''). These clusters are ranked by prevalence to produce the ordered lists presented in the results. Device and temporal breakdowns are computed as in our previous work \parencite{costagomes2025}, with mobile-to-desktop ratios for each health intent and hourly distributions across the day.

\vspace{-8pt}
\begin{center}
\begin{longtable}{>{\raggedright\arraybackslash}p{7cm} >{\raggedright\arraybackslash}p{7cm}}
\caption{Health Intent taxonomy.} \label{tab:intent} \\
\toprule
\textbf{Intent} & \textbf{Description} \\
\midrule
\endfirsthead
\toprule
\textbf{Intent} & \textbf{Description} \\
\midrule
\endhead
Health Information \& Education & General information about health and wellness topics \\
Symptom Questions \& Health Concerns & Questions about symptoms, health concerns, and understanding test results \\
Condition Information \& Care Questions & Questions about conditions, medications, and ongoing care \\
Fitness, Lifestyle \& Coaching & Nutrition, fitness, sleep, habit formation \\
Emotional Wellbeing & Coping strategies, stress management, behavioral health practices \\
Healthcare Navigation \& Access to Care & Finding care, appointments, care coordination \\
Coverage \& Benefits & Insurance, billing, cost, reimbursement \\
Research \& Academic Support & Literature review, exam preparation, curriculum development \\
Medical Paperwork & Medical documents, including forms, letters, or notes \\
Digital Tools \& Fitness Apps & Device setup, troubleshooting, data interpretation \\
Other Health / Fitness Intent & Health-related but uncategorized \\
Not Health & Primary intent unrelated to health \\
\bottomrule
\end{longtable}
\end{center}
\vspace{-8pt}
 
\section{Results}

\begin{figure}[H]
  \centering
  \includegraphics[width=\linewidth]{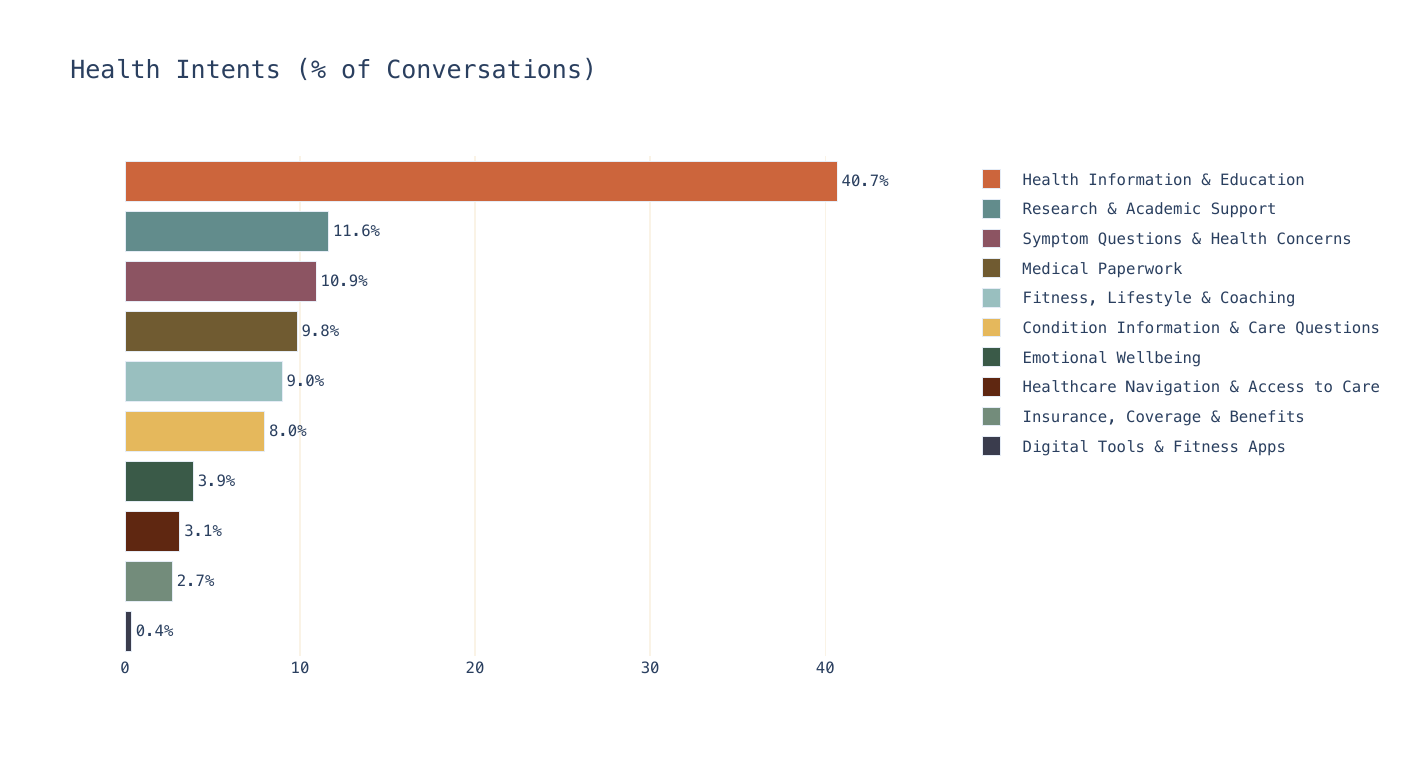}
  \caption{Distribution of health intent usage, in percentage of conversations, for the entire dataset.}
  \label{fig:intents_distribution}
\end{figure}

Figure \ref{fig:intents_distribution} shows the distribution of conversations across health intents. The largest category, ``Health Information \& Education,'' accounts for over 40\% of conversations. This category captures non-personalized queries, including how a medication works, what causes a condition, and general nutrition information. Its size is consistent with the finding that information seeking remains the dominant mode of health engagement online \parencite{eysenbach2002}. However, some queries framed in general terms may reflect underlying personal concerns, and the true share of personal health intents may be higher than the taxonomy suggests (see Section \ref{sec:limitations}). We observe from topic clusters (Section \ref{sec:results_topics}) that many queries are about specific treatments and conditions rather than general health education, further supporting our theory that users may seek general information as a step towards personal decision-making.

\begin{figure}[H]
  \centering
  \includegraphics[width=\linewidth]{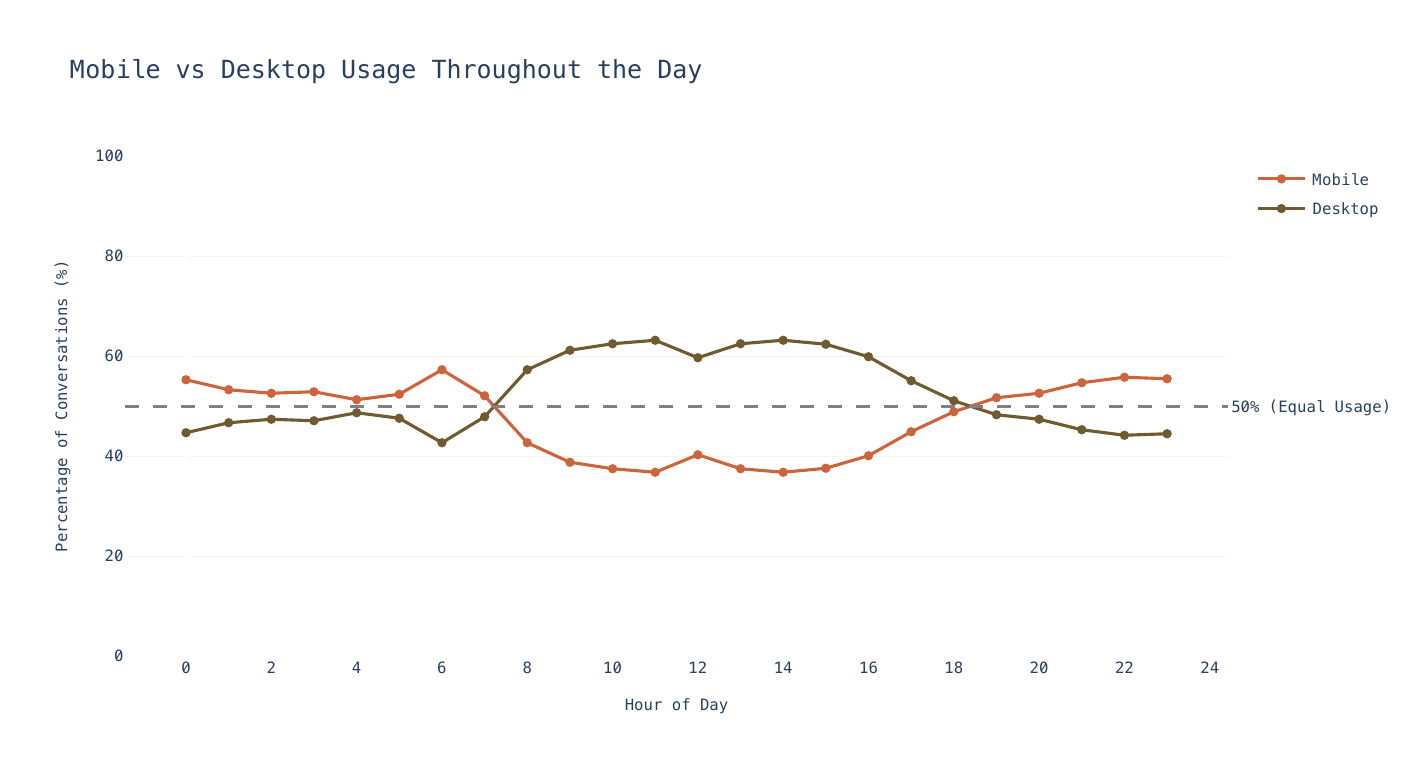}
  \caption{Average percentage of mobile vs desktop health conversations, throughout the day.}
  \label{fig:usage_mobile_vs_desktop}
\end{figure}

Figure \ref{fig:usage_mobile_vs_desktop} shows how the percentage of all conversations on desktop and mobile varies throughout the day, with the former more predominant during the day and the latter at night. This pattern reflects everyday routines: during working and school hours, users have access to desktop devices and may prefer them for longer or more complex tasks, whereas in the evening and at night, when people are away from their desks, the phone becomes the primary device for health queries.

\begin{figure}[H]
  \centering
  \includegraphics[width=\linewidth]{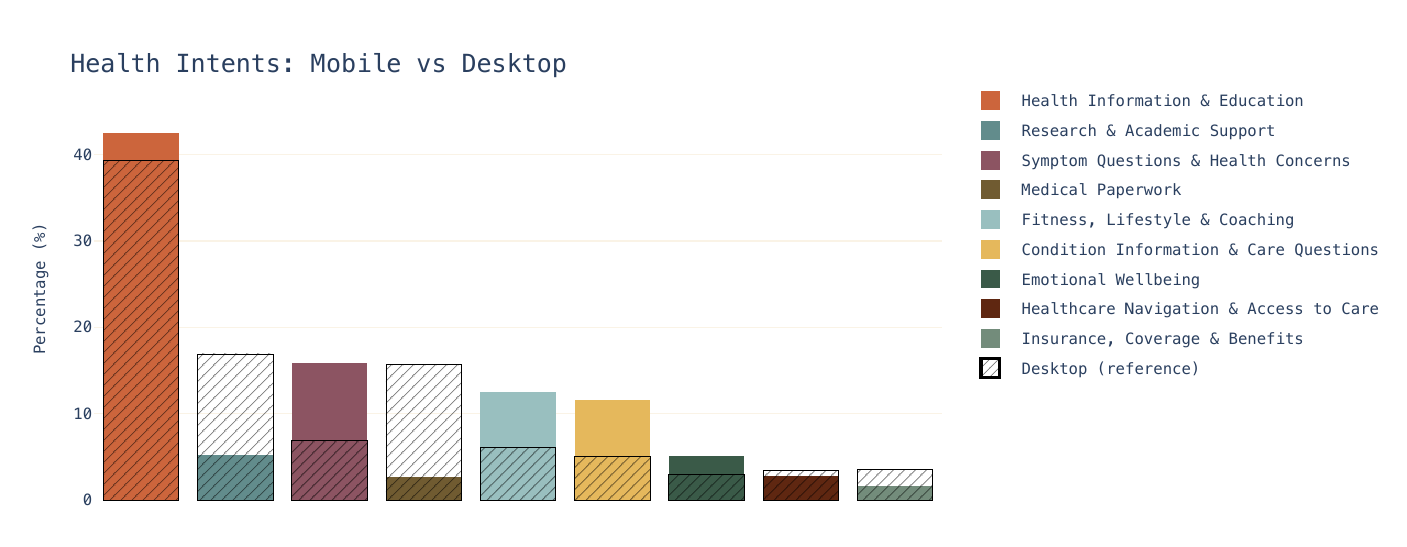}
  \caption{Average percentage of conversations per intent on mobile (block color) vs desktop (striped).}
  \label{fig:intents_desktop_vs_mobile}
\end{figure}

Figure \ref{fig:intents_desktop_vs_mobile} compares intent distributions across mobile and desktop. The most striking difference is in academic support and medical paperwork: these intents are among the most common on desktop but fall to the lower end on mobile. 

\begin{figure}[H]
  \centering
  \includegraphics[width=\linewidth]{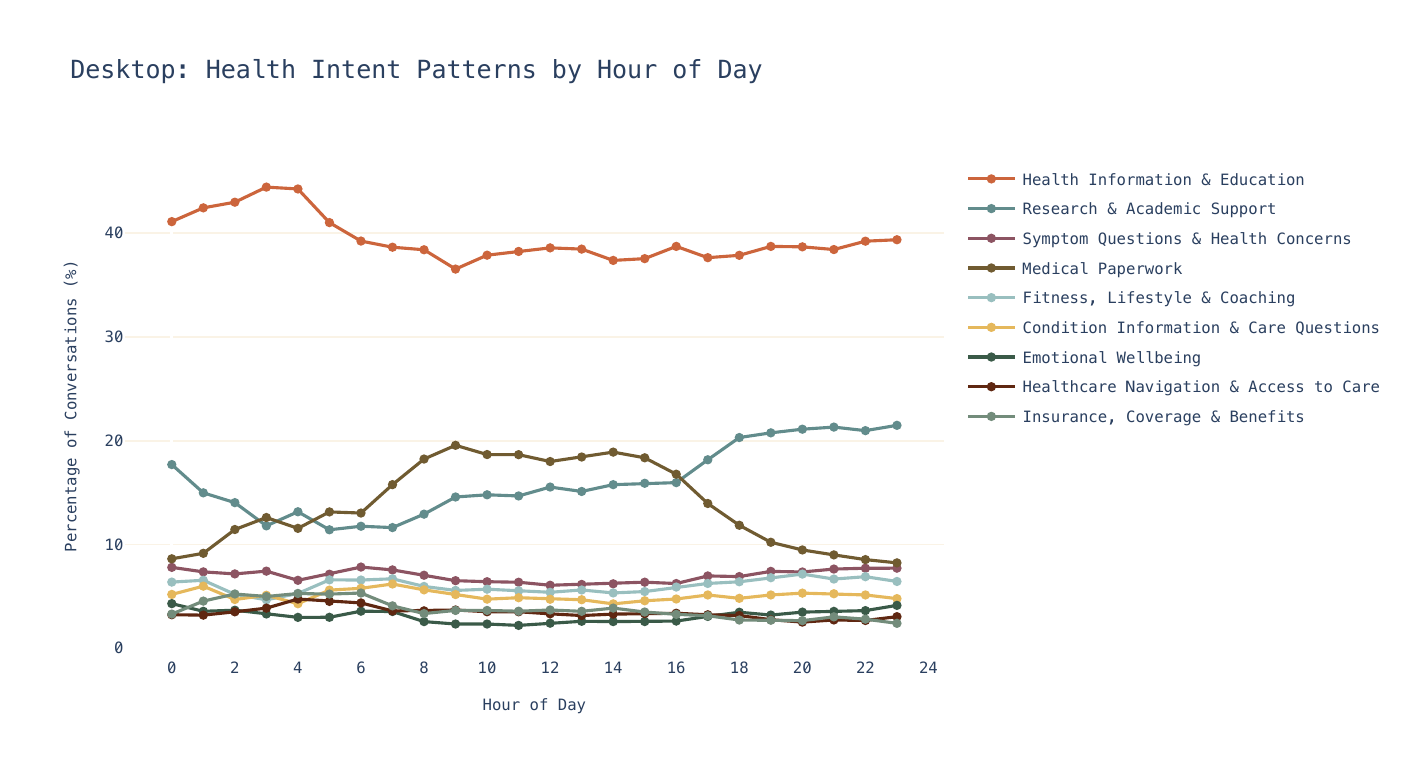}
  \caption{Health intent patterns averaged by hour of day, on desktop.}
  \label{fig:intent_time_of_day_desktop}
\end{figure}

\begin{figure}[H]
  \centering
  \includegraphics[width=\linewidth]{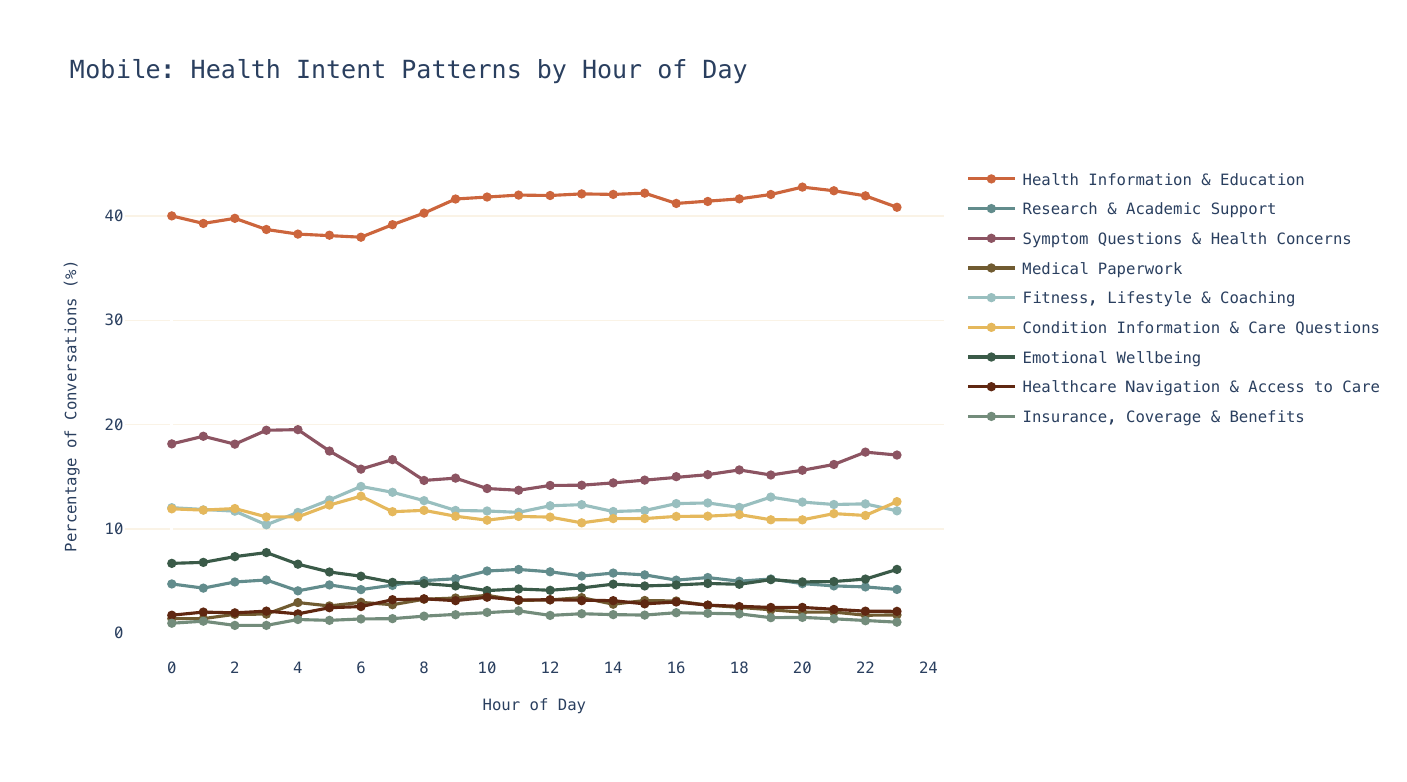}
  \caption{Health intent patterns averaged by hour of day, on mobile.}
  \label{fig:intent_time_of_day_mobile}
\end{figure}

Figures \ref{fig:intent_time_of_day_desktop} and \ref{fig:intent_time_of_day_mobile} show the breakdown of intents per hour of the day. Although the predominance of ``Health Information \& Education'' on both platforms is expected given Figure~\ref{fig:intents_desktop_vs_mobile}, on desktop its share decreases during working hours as ``Research \& Academic Support'' and ``Medical Paperwork'' rise. This suggests that Copilot usage on desktop is often adjacent to another activity such as thesis writing, research, or processing paperwork, tasks that typically require access to other documents or files alongside the conversation. ``Medical Paperwork'' peaks during normal working hours, while ``Research \& Academic Support'' rises steadily throughout the day, particularly after work and school hours when researchers and students turn to their own projects. More broadly, the desktop pattern may reflect workflows that depend on multiple windows and reference materials, which are cumbersome to manage on a mobile device.

On mobile, the second most common intent is ``Symptom Questions \& Health Concerns'', followed by queries on conditions and fitness. This is consistent with mobile devices being used primarily for personal health queries rather than work-related tasks.

\begin{figure}[H]
  \centering
  \includegraphics[width=0.95\linewidth]{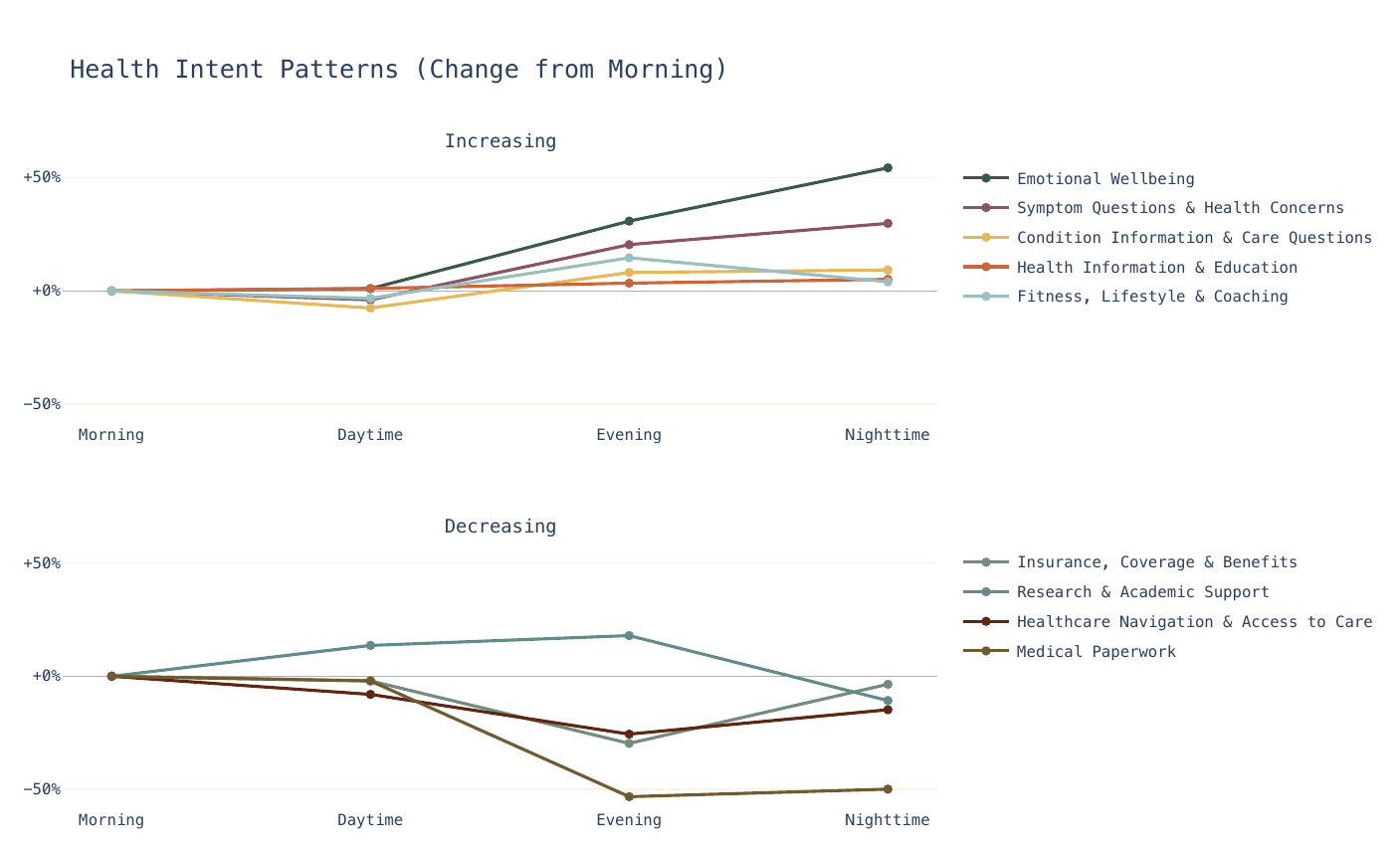}
  \caption{Temporal changes of intent usage, relative to the morning. The top graph shows the intents that increase throughout the day, and the bottom shows the ones that decrease. }
  \label{fig:intent_time_of_day_relative_to_morning}
\end{figure}

When looking at the changes throughout the day compared to morning, the distinction between types of intents becomes more evident (Figure \ref{fig:intent_time_of_day_relative_to_morning}), with the more personal intents (such as queries about conditions or emotional wellbeing) going up in the evening and at night and the more scholarly ones (such as research or documentation) decreasing.

\begin{figure}[H]
  \centering
  \includegraphics[width=0.95\linewidth]{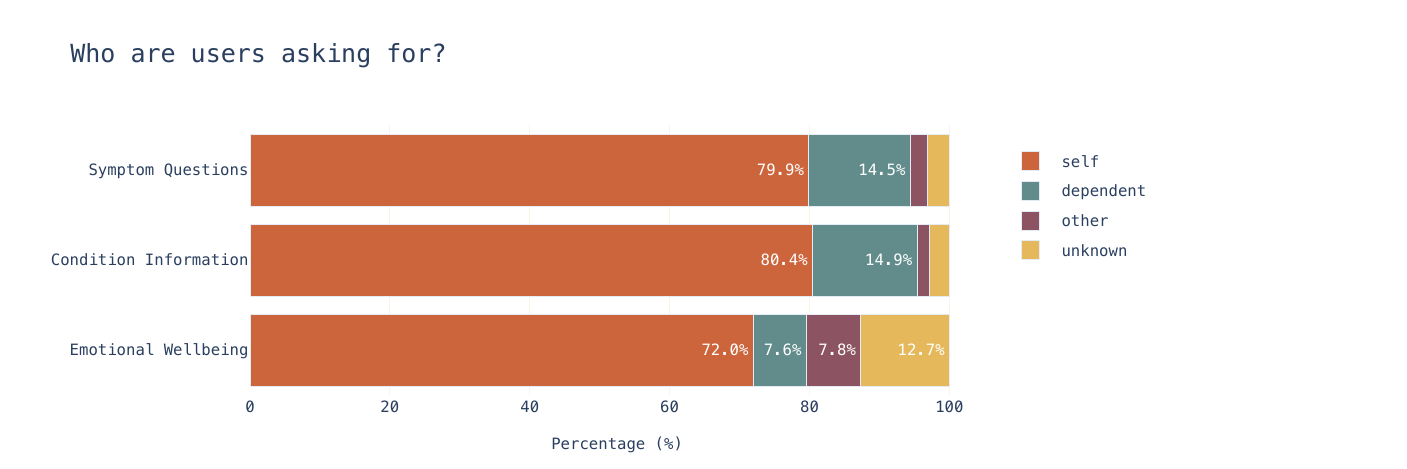}
  \caption{Percentage of conversations on three intents (symptom questions, condition information and emotional wellbeing) related to the user, a dependent, other or unknown.}
  \label{fig:personas}
\end{figure}

We also examined who the health query is about (Figure \ref{fig:personas}). In every category, most questions are asked on behalf of the users themselves. However, across both condition information and symptom questions, one in seven conversations are on behalf of someone else, whether a child, an aging parent, or a partner. 

\label{sec:results_topics} Table~\ref{tab:topics} presents the five most common topic clusters for six key intents, with within-category percentages. The clusters reveal that even the broadest category, ``Health Information \& Education'', is dominated by queries about specific treatments and conditions rather than abstract health knowledge, and that the narrower personal intents show clear concentrations around a small number of core needs.

\begin{center}
\begin{longtable}{>{\raggedright\arraybackslash}p{3.5cm} >{\raggedright\arraybackslash}p{9.5cm} >{\raggedright\arraybackslash}p{1.5cm}}
\caption{Top ranked topics of conversation per intent.} \label{tab:topics} \\
\toprule
\textbf{Intent} & \textbf{Top Topics} & \textbf{\%} \\
\midrule
\endfirsthead
\endhead
\vspace{2pt}%
Health Information \& Education & 
\begin{minipage}[t]{\linewidth}%
\vspace{2pt}%
\begin{itemize}[nosep,leftmargin=1.5em]
\item[1.] How specific treatments, medications, or medical procedures work
\item[2.] Causes, symptoms, and risk factors for diseases or conditions
\item[3.] Advice on healthy eating, supplements, and food safety
\item[4.] Plain‑language explanations of anatomy and biological concepts
\item[5.] Everyday wellness, prevention, and self‑care guidance
\end{itemize}  
\vspace{4pt}%
\end{minipage} &
\begin{minipage}[t]{\linewidth}%
\vspace{2pt}%
\begin{itemize}[nosep,leftmargin=0em,label={}]
\item 18.8
\item
\item 11.9
\item
\item 9.7
\item 8.0
\item
\item 6.5
\end{itemize}  
\vspace{4pt}%
\end{minipage}\\
\hline

\vspace{2pt}%
Symptom Questions \& Health Concerns & 
\begin{minipage}[t]{\linewidth}%
\vspace{2pt}%
\begin{itemize}[nosep,leftmargin=1.5em]
\item[1.] Understanding new or unexpected symptoms
\item[2.] Making sense of recurring or long‑term symptoms
\item[3.] Plain‑language explanations of lab or imaging results
\item[4.] Medication safety, side effects, and interactions
\item[5.] Infant/child health and development
\end{itemize} 
\vspace{4pt}%
\end{minipage} &
\begin{minipage}[t]{\linewidth}%
\vspace{2pt}%
\begin{itemize}[nosep,leftmargin=0em,label={}]
\item 32.7
\item 14.6
\item 12.8
\item 6.8
\item 6.8
\end{itemize} 
\vspace{4pt}%
\end{minipage}\\
\hline

\vspace{2pt}%
Condition Information \& Care Questions & 
\begin{minipage}[t]{\linewidth}%
\vspace{2pt}%
\begin{itemize}[nosep,leftmargin=1.5em]
\item[1.] Questions about medications, supplements, and treatments
\item[2.] Understanding how conditions develop and change over time
\item[3.] Caring for minor injuries at home
\item[4.] Everyday routines and self‑care for chronic conditions
\item[5.] Practical advice for skin and hair issues
\end{itemize}  
\vspace{4pt}%
\end{minipage} &
\begin{minipage}[t]{\linewidth}%
\vspace{2pt}%
\begin{itemize}[nosep,leftmargin=0em,label={}]
\item 18.5
\item
\item 12.2
\item
\item 12.1
\item 10.4
\item 7.3
\end{itemize}  
\vspace{4pt}%
\end{minipage}\\
\hline

\vspace{2pt}%
Fitness, Lifestyle \& Coaching & 
\begin{minipage}[t]{\linewidth}%
\vspace{2pt}%
\begin{itemize}[nosep,leftmargin=1.5em]
\item[1.] Tailored meal plans, calorie targets, and nutrition advice for specific goals
\item[2.] How to start or progress strength and resistance training
\item[3.] Support for setting and tracking fitness outcomes
\item[4.] Choosing healthier foods and tracking nutrition
\item[5.] Cardio, running, or cycling plans to build endurance
\end{itemize}  
\vspace{4pt}%
\end{minipage} &
\begin{minipage}[t]{\linewidth}%
\vspace{2pt}%
\begin{itemize}[nosep,leftmargin=0em,label={}]
\item 26.6
\item
\item 13.4
\item
\item 5.2
\item 5.1
\item 4.9
\end{itemize}  
\vspace{4pt}%
\end{minipage}\\
\hline
\vspace{2pt}%
Emotional Wellbeing & 
\begin{minipage}[t]{\linewidth}%
\vspace{2pt}%
\begin{itemize}[nosep,leftmargin=1.5em]
\item[1.] Understanding personal emotional or behavioral health challenges
\item[2.] Asking for practical routines to increase resilience and manage stress
\item[3.] Support for current emotional challenges
\item[4.] Supporting children, friends, or family members with emotional or behavioral health challenges
\item[5.] Social, academic, or workplace-related stress
\end{itemize} 
\vspace{4pt}%
\end{minipage} &
\begin{minipage}[t]{\linewidth}%
\vspace{2pt}%
\begin{itemize}[nosep,leftmargin=0em,label={}]
\item 30.5
\item
\item 13.4
\item
\item 12.2
\item 11.5
\item
\item 7.3
\end{itemize} 
\vspace{4pt}%
\end{minipage}\\

\hline

\vspace{2pt}%
Healthcare Navigation \& Access to Care & 
\begin{minipage}[t]{\linewidth}%
\vspace{2pt}%
\begin{itemize}[nosep,leftmargin=1.5em]
\item[1.] Finding nearby providers, clinics, or specialists
\item[2.] Help with medical paperwork and eligibility documentation
\item[3.] Comparing hospitals, clinics, procedures, and pricing
\item[4.] Understanding insurance coverage and benefits
\item[5.] Step‑by‑step help booking appointments
\end{itemize} 
\vspace{4pt}%
\end{minipage} &
\begin{minipage}[t]{\linewidth}%
\vspace{2pt}%
\begin{itemize}[nosep,leftmargin=0em,label={}]
\item 43.1
\item 11.5
\item
\item 7.7
\item 6.7
\item 6.4
\end{itemize} 
\vspace{4pt}%
\end{minipage}\\
\bottomrule
\end{longtable}
\end{center}

\section{Discussion}

The patterns that emerge from this analysis have implications for the design of health AI and for understanding health needs that existing systems may not be meeting.

\subsection{The after-hours pattern}

Personal health queries, particularly emotional wellbeing and symptom assessment, go up in the evening and at nighttime hours, precisely when traditional healthcare services are least accessible.

The emotional wellbeing pattern deserves particular attention. The evening increase in emotional health queries is consistent with the diurnal pattern of negative affect documented in population psychology, where negative affect rises throughout the day and peaks in the evening hours \parencite{golder2011}. This pattern is likely multiply determined: people may also have more time for personal reflection in the evening, and the reduced availability of professional support may itself prompt queries that would otherwise be directed to a clinician. The convergence between our observed intent patterns and an independently established affective rhythm provides suggestive evidence for the construct validity of our classification approach. 

\subsection{The boundary between general information and personal health queries}
Nearly one in five conversations involve users describing their own symptoms, interpreting their own test results, or managing their own conditions. These interactions exist on a spectrum: at one end, a user asking ``what does high cholesterol mean'' is seeking general education; at the other, a user describing persistent headaches alongside their medication list is seeking information specific to their own circumstances. Understanding the distribution and nature of these queries is a prerequisite for ensuring that conversational AI responses are appropriate and that users are directed to professional care when needed, which is what this taxonomy strives to enable.

\subsection{Device as a signal for intent}
Usage diverges sharply by device alongside previously shown lines \parencite{costagomes2025}. On mobile, personal health intents are substantially more prevalent, while desktop use is dominated by research, academic support, and medical paperwork. This split suggests that device choice is not merely a matter of convenience but reflects fundamentally different modes of health engagement. This distinction has practical implications for how health AI experiences are designed across platforms, and suggests that device context could inform how responses are prioritized and presented.

\subsection{Conversational AI as a health companion}

One in seven queries about symptoms and conditions are asked on behalf of someone else, e.g. a child, an aging parent, a partner. This finding reframes how we should think about health AI users. The person typing is not always the person the query is about. This also has design implications: a caregiver asking about an infant's or an elderly relative's symptoms may need different information, different contextual cues, and different follow-up recommendations than someone asking about their own. It has further safety implications: when a user is asking about a dependent, it may be for reasons that affect the accuracy and completeness of the information provided.

\subsection{Healthcare navigation as an unmet need}
The prevalence of queries including finding providers, understanding insurance, and completing paperwork, reveals that a meaningful fraction of health AI use addresses the complexity of healthcare systems rather than health itself. Users are asking AI to help them do things that should, in principle, be straightforward: find a doctor, book an appointment, understand what their insurance covers. The fact that these queries exist at such volume provides a signal about friction in existing healthcare delivery.

\subsection{Limitations}
\label{sec:limitations}
This study has several important limitations. First, our analysis was conducted exclusively on Microsoft Copilot consumer logs, representing a specific user population and platform context. While our findings directly inform design and safety considerations for this platform and similar general-purpose AI assistants, generalization to other platforms, clinical settings, or populations may be limited. Second, we observe queries but not outcomes: we cannot determine whether users subsequently sought clinical care, how they interpreted responses, or whether the information they received improved their health decisions. Third, our sample is drawn from a single month (January 2026), and seasonal effects may influence the distribution of intents. January, in particular, is associated with New Year's health resolutions, which may influence fitness and lifestyle queries relative to other months. Fourth, our taxonomy captures intent as expressed in conversation, not the underlying clinical need. The ``Health Information \& Education'' category accounts for over 40\% of conversations, and while its topic clusters suggest meaningful heterogeneity within it, its size may partly reflect the inherent difficulty of distinguishing general from personal information seeking. Conversations do not always contain sufficient context to determine whether a generally framed query (e.g., ``what are the side effects of metformin'') reflects casual curiosity or a user's own medication concern. This means our classifier defaults to the less-specific educational label in ambiguous cases, and the reported share of personal health intents may represent a lower bound. Future iterations of the taxonomy should explore whether this category can be further subdivided.

\subsection{Future directions}
This work opens several future directions. Longitudinally, tracking how intent distributions shift as conversational AI matures will reveal whether users discover new applications or converge on established patterns. Geographically, understanding how health AI usage differs across regions and healthcare systems, particularly between settings with strong primary care access and those without, will be essential for responsible global deployment. Methodologically, linking intents to response quality and downstream outcomes would move the field from characterizing what people ask to evaluating whether what they receive helps them. And from a safety perspective, the personal health intents identified here such as symptom assessment, condition management, and emotional wellbeing, arguably define some of the categories where the consequences of conversational AI responses are highest, and where investment in response quality and safety measures should be concentrated.

\pagebreak

\printbibliography

@misc{costagomes2025,
      title={It's About Time: The Temporal and Modal Dynamics of Copilot Usage}, 
      author={Beatriz Costa-Gomes and Sophia Chen and Connie Hsueh and Deborah Morgan and Philipp Schoenegger and Yash Shah and Sam Way and Yuki Zhu and Timothé Adeline and Michael Bhaskar and Mustafa Suleyman and Seth Spielman},
      year={2025},
      eprint={2512.11879},
      archivePrefix={arXiv},
      primaryClass={cs.CY},
      url={https://arxiv.org/abs/2512.11879}, 
}

@article {eysenbach2002,
	author = {Eysenbach, Gunther and K{\"o}hler, Christian},
	title = {How do consumers search for and appraise health information on the world wide web? Qualitative study using focus groups, usability tests, and in-depth interviews},
	volume = {324},
	number = {7337},
	pages = {573--577},
	year = {2002},
	doi = {10.1136/bmj.324.7337.573},
	publisher = {BMJ Publishing Group Ltd},
	issn = {0959-8138},
	URL = {https://www.bmj.com/content/324/7337/573},
	eprint = {https://www.bmj.com/content/324/7337/573.full.pdf},
	journal = {BMJ}
}

@online{anthropic2025,
author = {Miles McCain and Ryn Linthicum and Chloe Lubinski and Alex Tamkin and Saffron Huang and Michael Stern and Kunal Handa and Esin Durmus and Tyler Neylon and Stuart Ritchie and Kamya Jagadish and Paruul Maheshwary and Sarah Heck and Alexandra Sanderford and Deep Ganguli},
title = {How People Use Claude for Support, Advice, and Companionship},
date = {2025-06-26},
year = {2025},
url = {https://www.anthropic.com/news/how-people-use-claude-for-support-advice-and-companionship},
}

@techreport{openai2025,
 title = "How People Use ChatGPT",
 author = "Chatterji, Aaron and Cunningham, Thomas and Deming, David J and Hitzig, Zoe and Ong, Christopher and Shan, Carl Yan and Wadman, Kevin",
 institution = "National Bureau of Economic Research",
 type = "Working Paper",
 series = "Working Paper Series",
 number = "34255",
 year = "2025",
 month = "09",
 doi = {10.3386/w34255},
 URL = "http://www.nber.org/papers/w34255"
}

@article{kung2023,
    doi = {10.1371/journal.pdig.0000198},
    author = {Kung, Tiffany H. AND Cheatham, Morgan AND Medenilla, Arielle AND Sillos, Czarina AND De Leon, Lorie AND Elepaño, Camille AND Madriaga, Maria AND Aggabao, Rimel AND Diaz-Candido, Giezel AND Maningo, James AND Tseng, Victor},
    journal = {PLOS Digital Health},
    publisher = {Public Library of Science},
    title = {Performance of ChatGPT on USMLE: Potential for AI-assisted medical education using large language models},
    year = {2023},
    month = {02},
    volume = {2},
    url = {https://doi.org/10.1371/journal.pdig.0000198},
    pages = {1-12},
    number = {2},

}

@Article{tan2017,
author="Tan, Sharon Swee-Lin
and Goonawardene, Nadee",
title="Internet Health Information Seeking and the Patient-Physician Relationship: A Systematic Review",
journal="J Med Internet Res",
year="2017",
month="01",
day="19",
volume="19",
number="1",
pages="e9",
issn="1438-8871",
doi="10.2196/jmir.5729",
url="http://www.jmir.org/2017/1/e9/",
url="https://doi.org/10.2196/jmir.5729",
url="http://www.ncbi.nlm.nih.gov/pubmed/28104579"
}

@article{lee2023,
author = {Peter Lee  and Sebastien Bubeck  and Joseph Petro },
title = {Benefits, Limits, and Risks of GPT-4 as an AI Chatbot for Medicine},
journal = {New England Journal of Medicine},
volume = {388},
number = {13},
pages = {1233-1239},
year = {2023},
doi = {10.1056/NEJMsr2214184},

URL = {https://www.nejm.org/doi/full/10.1056/NEJMsr2214184},
eprint = {https://www.nejm.org/doi/pdf/10.1056/NEJMsr2214184}
}

@Article{Singhal2023,
author={Singhal, Karan and Azizi, Shekoofeh and Tu, Tao and Mahdavi, S. Sara and Wei, Jason and Chung, Hyung Won and Scales, Nathan and Tanwani, Ajay and Cole-Lewis, Heather and Pfohl, Stephen and Payne, Perry and Seneviratne, Martin and Gamble, Paul and Kelly, Chris and Babiker, Abubakr and Sch{\"a}rli, Nathanael and Chowdhery, Aakanksha and Mansfield, Philip and Demner-Fushman, Dina and Ag{\"u}era y Arcas, Blaise and Webster, Dale and Corrado, Greg S. and Matias, Yossi and Chou, Katherine and Gottweis, Juraj and Tomasev, Nenad and Liu, Yun and Rajkomar, Alvin and Barral, Joelle and Semturs, Christopher and Karthikesalingam, Alan and Natarajan, Vivek},
title={Large language models encode clinical knowledge},
journal={Nature},
year={2023},
month={08},
day={01},
volume={620},
number={7972},
pages={172-180},
issn={1476-4687},
doi={10.1038/s41586-023-06291-2},
url={https://doi.org/10.1038/s41586-023-06291-2}
}

@inproceedings{wan2024,
  title={Tnt-llm: Text mining at scale with large language models},
  author={Wan, Mengting and Safavi, Tara and Jauhar, Sujay Kumar and Kim, Yujin and Counts, Scott and Neville, Jennifer and Suri, Siddharth and Shah, Chirag and White, Ryen W and Yang, Longqi and others},
  booktitle={Proceedings of the 30th ACM SIGKDD conference on knowledge discovery and data mining},
  pages={5836--5847},
  year={2024}
}

@article{
golder2011,
author = {Scott A. Golder  and Michael W. Macy },
title = {Diurnal and Seasonal Mood Vary with Work, Sleep, and Daylength Across Diverse Cultures},
journal = {Science},
volume = {333},
number = {6051},
pages = {1878-1881},
year = {2011},
doi = {10.1126/science.1202775},
URL = {https://www.science.org/doi/abs/10.1126/science.1202775},
eprint = {https://www.science.org/doi/pdf/10.1126/science.1202775},

}

@article{nori2024medprompto1explorationruntime,
      title={From Medprompt to o1: Exploration of Run-Time Strategies for Medical Challenge Problems and Beyond}, 
      author={Harsha Nori and Naoto Usuyama and Nicholas King and Scott Mayer McKinney and Xavier Fernandes and Sheng Zhang and Eric Horvitz},
      year={2024},
      eprint={2411.03590},
      archivePrefix={arXiv},
      primaryClass={cs.CL},
      url={https://arxiv.org/abs/2411.03590}, 
}

@article{nori2023capabilitiesgpt4medicalchallenge,
      title={Capabilities of GPT-4 on Medical Challenge Problems}, 
      author={Harsha Nori and Nicholas King and Scott Mayer McKinney and Dean Carignan and Eric Horvitz},
      year={2023},
      eprint={2303.13375},
      archivePrefix={arXiv},
      primaryClass={cs.CL},
      url={https://arxiv.org/abs/2303.13375}, 
}

@article{nori2025sequentialdiagnosislanguagemodels,
      title={Sequential Diagnosis with Language Models}, 
      author={Harsha Nori and Mayank Daswani and Christopher Kelly and Scott Lundberg and Marco Tulio Ribeiro and Marc Wilson and Xiaoxuan Liu and Viknesh Sounderajah and Jonathan Carlson and Matthew P Lungren and Bay Gross and Peter Hames and Mustafa Suleyman and Dominic King and Eric Horvitz},
      year={2025},
      eprint={2506.22405},
      archivePrefix={arXiv},
      primaryClass={cs.CL},
      url={https://arxiv.org/abs/2506.22405}, 
}

@misc{brodeur2025superhumanperformancelargelanguage,
      title={Superhuman performance of a large language model on the reasoning tasks of a physician}, 
      author={Peter G. Brodeur and Thomas A. Buckley and Zahir Kanjee and Ethan Goh and Evelyn Bin Ling and Priyank Jain and Stephanie Cabral and Raja-Elie Abdulnour and Adrian D. Haimovich and Jason A. Freed and Andrew Olson and Daniel J. Morgan and Jason Hom and Robert Gallo and Liam G. McCoy and Haadi Mombini and Christopher Lucas and Misha Fotoohi and Matthew Gwiazdon and Daniele Restifo and Daniel Restrepo and Eric Horvitz and Jonathan Chen and Arjun K. Manrai and Adam Rodman},
      year={2025},
      eprint={2412.10849},
      archivePrefix={arXiv},
      primaryClass={cs.AI},
      url={https://arxiv.org/abs/2412.10849}, 
}

@Article{topol2019,
author={Topol, Eric J.},
title={High-performance medicine: the convergence of human and artificial intelligence},
journal={Nature Medicine},
year={2019},
month={01},
day={01},
volume={25},
number={1},
pages={44-56},
issn={1546-170X},
doi={10.1038/s41591-018-0300-7},
url={https://doi.org/10.1038/s41591-018-0300-7}
}

@article{huo2025large,
  title={Large language models for chatbot health advice studies: a systematic review},
  author={Huo, Bright and Boyle, Amy and Marfo, Nana and Tangamornsuksan, Wimonchat and Steen, Jeremy P and McKechnie, Tyler and Lee, Yung and Mayol, Julio and Antoniou, Stavros A and Thirunavukarasu, Arun James and others},
  journal={JAMA Network Open},
  volume={8},
  number={2},
  pages={e2457879},
  year={2025}
}

@misc{goldberg2024no,
  title={To do no harm—and the most good—with AI in health care},
  author={Goldberg, Carey Beth and Adams, Laura and Blumenthal, David and Brennan, Patricia Flatley and Brown, Noah and Butte, Atul J and Cheatham, Morgan and DeBronkart, Dave and Dixon, Jennifer and Drazen, Jeff and others},
  journal={Nejm Ai},
  volume={1},
  number={3},
  pages={AIp2400036},
  year={2024},
  publisher={Massachusetts Medical Society}
}

@article{ruben2025artificial,
  title={What is Artificial Intelligence (AI)“Empathy”? A Study Comparing ChatGPT and Physician Responses on an Online Forum: Ruben et al.},
  author={Ruben, Mollie A and Blanch-Hartigan, Danielle and Hall, Judith A},
  journal={Journal of General Internal Medicine},
  pages={1--8},
  year={2025},
  publisher={Springer}
}

@article{lizee2024conversational,
  title={Conversational medical AI: ready for practice},
  author={Liz{\'e}e, Antoine and Beaucot{\'e}, Pierre-Auguste and Whitbeck, James and Doumeingts, Marion and Beaugnon, Ana{\"e}l and Feldhaus, Isabelle},
  journal={arXiv preprint arXiv:2411.12808},
  year={2024}
}

@article{ramaswamy2026chatgpt,
  title={ChatGPT Health performance in a structured test of triage recommendations},
  author={Ramaswamy, Ashwin and Tyagi, Alvira and Hugo, Hannah and Jiang, Joy and Jayaraman, Pushkala and Jangda, Mateen and Te, Alexis E and Kaplan, Steven A and Lampert, Joshua and Freeman, Robert and others},
  journal={Nature Medicine},
  year={2026},
  publisher={Nature Publishing Group}
}

@article{bean2026reliability,
  title={Reliability of LLMs as medical assistants for the general public: a randomized preregistered study},
  author={Bean, Andrew M and Payne, Rebecca Elizabeth and Parsons, Guy and Kirk, Hannah Rose and Ciro, Juan and Mosquera-G{\'o}mez, Rafael and Hincapi{\'e} M, Sara and Ekanayaka, Aruna S and Tarassenko, Lionel and Rocher, Luc and others},
  journal={Nature Medicine},
  pages={1--7},
  year={2026},
  publisher={Nature Publishing Group US New York}
}

\end{document}